\documentstyle[12pt,prd,aps,epsfig,preprint]{revtex}

\begin{document}
\title{{\bf Exact Solution of the Klein-Gordon Equation for the }${\cal PT}${\bf %
-Symmetric Generalized Woods-Saxon Potential by the Nikiforov-Uvarov Method }}
\author{Sameer M. Ikhdair\thanks{%
sikhdair@neu.edu.tr} and \ Ramazan Sever\thanks{%
sever@metu.edu.tr}}
\address{$^{\ast }$Department of Physics, \ Near East University, Nicosia, North
Cyprus, Mersin 10, Turkey\\
$^{\dagger }$Department of Physics, Middle East Technical University, 06531
Ankara, Turkey.}
\date{\today
}
\maketitle

\begin{abstract}
The one-dimensional Klein-Gordon (KG) equation has been solved for the $%
{\cal PT}$-symmetric generalized Woods-Saxon (WS) potential. The
Nikiforov-Uvarov (${\rm NU}$) method which is based on solving the
second-order linear differential equations by reduction to a generalized
equation of hypergeometric type is used to obtain exact energy eigenvalues
and corresponding eigenfunctions. We have also investigated the positive and
negative exact bound states of the ${\rm s}$-states for different types of
complex generalized WS potentials.


\end{abstract}


\section{Introduction}

\noindent In the past few years there has been considerable work on
non-Hermitian Hamiltonians. Among this kind of Hamiltonians, much attention
has been focused on the investigation of properties of so-called ${\cal PT}$%
-symmetric Hamiltonians. Following the early studies of Bender {\it et al}.
[1], the ${\cal PT}$-symmetry formulation has been successfuly utilized by
many authors [2-8]. The ${\cal PT}$-symmetric but non-Hermitian Hamiltonians
have real spectra whether the Hamiltonians are Hermitian or not.
Non-Hermitian Hamiltonians with real or complex spectra have also been
analyzed by using different methods [3-6,9]. Non-Hermitian but ${\cal PT}$%
-symmetric models have applications in different fields, such as optics
[10], nuclear physics [11], condensed matter [12], quantum field theory [13]
and population biology [14].

Exact solution of Schr\"{o}dinger equation for central potentials has
generated much interest in recent years. So far, some of these potentials
are the parabolic type potential [15], the Eckart potential [16,17], the
Fermi-step potential [16,17], the Rosen-Morse potential [18], the Ginocchio
barrier [19], the Scarf barriers [20], the Morse potential [21] and a
potential which interpolates between Morse and Eckart barriers [22]. Many
authors have studied on exponential type potentials [23-26] and quasi
exactly solvable quadratic potentials [27-29]. In addition, Schr\"{o}dinger,
Dirac, Klein-Gordon, and Duffin-Kemmer-Petiau equations for a Coulomb type
potential are solved by using different method [30-34]. The exact solutions
for these models have been obtained analytically.

Further, using the quantization of the boundary condition of the states at
the origin, Znojil [35] studied another generalized Hulth\'{e}n and other
exponential potentials in non-relativistic and relativistic regions.
Domingues-Adame [36] and Chetouani {\it et al.} [37] also studied
relativistic bound states of the standard Hulth\'{e}n potential. On the
other hand, Rao and Kagali [38] investigated the relativistic bound states
of the exponential-type screened Coulomb potential by means of the
one-dimensional ($1D$) Klein-Gordon equation. However, it is well-known that
for the exponential-type screened Coulomb potential there is no explicit
form of the energy expression of bound-states for Schr\"{o}dinger [39], KG
[38] and also Dirac [16] equations. \c{S}im\c{s}ek and E\u{g}rifes [31] have
presented the bound-state solutions of $1D$ Klein-Gordon (KG) equation for $%
{\cal PT}$-symmetric potentials with complex generalized Hulth\'{e}n
potential. In a latter study, E\u{g}rifes and Sever [32] investigated the
bound-state solutions of the $1D$ Dirac equation with ${\cal PT}$-symmetric
and non-${\cal PT}$-symmetric real and complex forms of the generalized
Hulth\'{e}n potential. Yi {\it et al}. [40] obtained the energy equations in
the KG theory with equally mixed vector and scalar Rosen-Morse-type
potentials. Berkdemir {\it et al}. [41] obtained the bound-state eigenvalues
and eigenfunctions for the Schr\"{o}dinger equation using a generalized form
of Woods-Saxon (WS) potential by means of Nikoforov-Uvarov method [42]. In
recent works, we have investigated the $\ell \neq 0$ bound-state solutions
of the $1D$ Schr\"{o}dinger equation with the real and complex forms of the
modified Hulth\'{e}n and WS potentials [43,44] for their energy spectra and
correponding wave functions using the Nikiforov-Uvarov (NU) method. In
addition, we have extended our study to relativistic models in solving the
spinless Salpeter equation analytically for its exact bound-state spectra
and wavefunctions for the real and complex forms of the ${\cal PT}$%
-symmetric generalized Hulth\'{e}n potential [45].

In the present work, we investigate the bound-state solutions of the $1D$
Klein-Gordon (KG) equation with real and complex forms of the generalized
Woods-Saxon (WS) potential. Hence, the objective of our work is to determine
the exact energy levels of relativistic KG \ particles trapped in a
spherically symmetric generalized WS potential which possesses a ${\cal PT}$%
-symmetry as well. In this context, it is possible to convert KG to a
Schr\"{o}dinger-like equation which takes the hypergeometric form to which
one may apply the NU method.

The organization of this work is as follows. After a brief introductory
discussion of the NU method in Section \ref{TNU}, we discuss the KG problem
and then obtain the exact bound-state energy spectra for real and complex
cases of generalized WS potentials and their corresponding eigenfunctions in
Section \ref{KGP}. In Section \ref{RAC}, we discuss our results and mention
briefly the scope of this relativistic study.

\section{The Nikiforov-Uvarov Method}

\label{TNU}In this section we outline the basic formulations of the method.
The Schr\"{o}dinger equation and other Schr\"{o}dinger-type equations can be
solved by using the Nikiforov-Uvarov (${\rm NU}$) method which is based on
the solutions of general second-order linear differential equation with
special orthogonal functions [42]. It is well known that for any given $1D$
radial potential, the Schr\"{o}dinger equation can be reduced to a
generalized equation of hypergeometric type with an appropriate
transformation and it can be written in the following form

\begin{equation}
\psi _{n}^{\prime \prime }(s)+\frac{\widetilde{\tau }(s)}{\sigma (s)}\psi
_{n}^{\prime }(s)+\frac{\widetilde{\sigma }(s)}{\sigma ^{2}(s)}\psi
_{n}(s)=0,
\end{equation}
where $\sigma (s)$ and $\widetilde{\sigma }(s)$ are polynomials, at most of
second-degree, and $\widetilde{\tau }(s)$ is of a first-degree polynomial.
To find particular solution of Eq.(1) we apply the method of separation of
variables using the transformation

\begin{equation}
\psi _{n}(s)=\phi _{n}(s)y_{n}(s),
\end{equation}
which reduces Eq.(1) into a hypergeometric-type equation

\begin{equation}
\sigma (s)y_{n}^{\prime \prime }(s)+\tau (s)y_{n}^{\prime }(s)+\lambda
y_{n}(s)=0,
\end{equation}
whose polynomial solutions $y_{n}(s)$ of the hypergeometric type function
are given by Rodrigues relation

\begin{equation}
y_{n}(s)=\frac{B_{n}}{\rho (s)}\frac{d^{n}}{ds^{n}}\left[ \sigma ^{n}(s)\rho
(s)\right] ,\text{ \ \ \ }\left( n=0,1,2,...\right)
\end{equation}
where $B_{n}$ is a normalizing constant and $\rho (s)$ is the weight
function satisfying the condition [42]

\begin{equation}
\frac{d}{ds}w(s)=\frac{\tau (s)}{\sigma (s)}w(s),
\end{equation}
where $w(s)=\sigma (s)\rho (s).$ On the other hand, the function $\phi (s)$
satisfies the condition

\begin{equation}
\frac{d}{ds}\phi (s)=\frac{\pi (s)}{\sigma (s)}\phi (s),
\end{equation}
where the linear polynomial $\pi (s)$ is given by
\begin{equation}
\pi (s)=\frac{\sigma ^{\prime }(s)-\widetilde{\tau }(s)}{2}\pm \sqrt{\left(
\frac{\sigma ^{\prime }(s)-\widetilde{\tau }(s)}{2}\right) ^{2}-\widetilde{%
\sigma }(s)+k\sigma (s)},
\end{equation}
from which the root $k$ is the essential point in the calculation of $\pi
(s) $ is determined$.$ Further, the parameter $\lambda $ required for this
method is defined as
\begin{equation}
\lambda =k+\pi ^{\prime }(s).
\end{equation}
Further, in order to find the value of $k,$ the discriminant under the
square root is being set equal to zero and the resulting second-order
polynomial has to be solved for its roots $k_{+,-}$. Thus, a new eigenvalue
equation for the ${\rm SE}$ becomes

\begin{equation}
\lambda _{n}+n\tau ^{\prime }(s)+\frac{n\left( n-1\right) }{2}\sigma
^{\prime \prime }(s)=0,\text{ \ \ \ \ \ \ }\left( n=0,1,2,...\right)
\end{equation}
where

\begin{equation}
\tau (s)=\widetilde{\tau }(s)+2\pi (s),
\end{equation}
and it must have a negative derivative.

\section{Exact Bound State Solutions of the Generalized Woods-Saxon Potential%
}

\label{KGP}In this section, we formulate the Nikiforov-Uvarov solution for
the relativistic motion of a spin-zero particle bound in spherically
symmetric well-known spherical symmetric generalized Woods-Saxon potential
of the form [46]
\begin{equation}
V_{q}(r,R_{0})=-V_{0}\frac{e^{-\left( \frac{r-R_{0}}{a}\right) }}{%
1+qe^{-\left( \frac{r-R_{0}}{a}\right) }},\text{ \ }q\geq 0,\text{ }R_{0}\gg
a,
\end{equation}
where $r$ refers to the center-of-mass distance between the projectile and
the target nuclei (the $r$ range is from $0$ to $\infty $). The relevant
parameters of the nuclear potential are given as follows: $%
R_{0}=r_{0}A^{1/3} $ is to define the confinement barrier position value of
the corresponding spherical nucleus or the width of the potential, $A$ is
the target mass number, $r_{0}$ is the radius parameter, $V_{0}$ controls
the barrier height of the Coulombic part, $a$ is the surface diffuseness
parameter has to control its slope, which is usually adjusted to the
experimental values of ionization energies. Further, $q$ is a shape
(deformation) parameter, the strength of the exponential part other than
unity, set to determine the shape of potential and is arbitrarily taken to
be a real constant within the potential. Further, we remark that the spatial
coordinates in the potential are not deformed and thus the potential still
remains spherical. Obviously, for some specific $q$ values this potential
reduces to the well-known types, such as for $q=0$ to the exponential
potential and for $q=-1$ and $a=\delta ^{-1}$ to the modified Hulth\'{e}n
potential (cf. [43,44,45] and the references therein).

For a scalar particle of rest mass $m$ and total energy $E_{nl},$ the radial
part of the KG equationin in three-dimensional spherical coordinates [47]

\begin{equation}
\left[ -\frac{\hbar ^{2}}{2m}\frac{d^{2}}{dr^{2}}+\frac{\hbar ^{2}l\left(
l+1\right) }{2mr^{2}}+\frac{1}{2mc^{2}}\left[ m^{2}c^{4}-\left[ E_{nl}-V(r)%
\right] ^{2}\right] \right] \psi _{nl}(r)=0,
\end{equation}
where $\psi _{nl}(r)$ is the reduced radial wave function.

To calculate the energy eigenvalues and the corresponding eigenfunctions, we
substitute the Hermitian real-valued $1D$ generalized WS potential (the $r$
range is from $0$ to $\infty $):

\begin{equation}
V_{q}(x)=-V_{0}\frac{e^{-\alpha x}}{1+qe^{-\alpha x}},\text{ \ }\alpha =1/a,%
\text{ }x=r-R_{0}
\end{equation}
into the $1D$ ${\cal PT}{\rm -}$symmetrical Hermitian spinless KG equation
(12) for $l=0$ case $($i.e., $s$-wave states$)$ and then obtain

\begin{equation}
\frac{d^{2}\psi _{nq}(x)}{dx^{2}}+\frac{1}{\hbar ^{2}c^{2}}\left[
(E_{n}^{2}-m^{2}c^{4})+V_{0}^{2}\frac{e^{-2\alpha x}}{(1+qe^{-\alpha x})^{2}}%
+2V_{0}E_{n}\frac{e^{-\alpha x}}{\left( 1+qe^{-\alpha x}\right) }\right]
\psi _{nq}(x)=0.
\end{equation}
We employ the following novel dimensionless transformation parameter, s$%
(x)=(1+qe^{-\alpha x})^{-1}$ , which is in real phase and found to maintain
the transformed wavefunctions finite on the boundary conditions (i.e., $%
-\infty \leq x\leq \infty \rightarrow 0\leq s\leq 1)$ [44]$.$ Hence,
applying such transformation to the upper equation after setting $\hbar
=c=1, $ it gives

\begin{equation}
\frac{d^{2}\psi _{nq}(s)}{ds^{2}}+\frac{1-2s}{s-s^{2}}\frac{d\psi _{nq}(s)}{%
ds}+\frac{1}{\alpha ^{2}(s-s^{2})^{2}}\left[ (E_{n}^{2}-m^{2})+\widetilde{V}%
_{0}^{2}(1-s)^{2}+2E_{n}\widetilde{V}_{0}(1-s)\right] \psi _{nq}(s)=0,
\end{equation}
where $\widetilde{V}_{0}=V_{0}/q.$ \ With the dimensionless definitions
given by
\begin{equation}
-\epsilon ^{2}=\frac{\left( E_{n}^{2}-m^{2}\right) }{\alpha ^{2}}\geq 0,%
\text{ }\beta ^{2}=\frac{2E_{n}\widetilde{V}_{0}}{\alpha ^{2}},\text{\ }%
\gamma ^{2}=\frac{\widetilde{V}_{0}^{2}}{\alpha ^{2}}\text{, (}E_{n}^{2}\leq
m^{2},\text{ }\beta ^{2}>0,\text{ }\gamma ^{2}>0\text{ }),
\end{equation}
for bound-states, one can arrive at the simple hypergeometric equation given
by

\begin{equation}
\psi {}_{nq}^{\prime \prime }(s)+\frac{1-2s}{(s-s^{2})}\psi {}_{nq}^{\prime
}(s)+\frac{\left[ s^{2}\gamma ^{2}-s\left( \beta ^{2}+2\gamma ^{2}\right) +%
\text{\ }\beta ^{2}+\gamma ^{2}-\epsilon ^{2}\right] }{\left( s-s^{2}\right)
^{2}}\psi _{nq}(s)=0.
\end{equation}
Hence, comparing the last equation with the generalized hypergeometric type,
Eq.(1), we obtain the associated polynomials as

\begin{equation}
\widetilde{\tau }(s)=1-2s,\text{ \ \ \ }\sigma (s)=(s-s^{2}),\text{ \ \ }%
\widetilde{\sigma }(s)=s^{2}\gamma ^{2}-s\left( \beta ^{2}+2\gamma
^{2}\right) +\text{\ }\beta ^{2}+\gamma ^{2}-\epsilon ^{2}.
\end{equation}
When these polynomials are substituted into Eq.(7), with $\sigma ^{\prime
}(s)=1-2s,$ we obtain

\begin{equation}
\pi (s)=\pm \sqrt{-s^{2}(\gamma ^{2}+k)+s\left( \beta ^{2}+2\gamma
^{2}+k\right) +\epsilon ^{2}-\text{\ }\beta ^{2}-\gamma ^{2}}.
\end{equation}
Further, the discriminant of the upper expression under the square root has
to be set equal to zero. Therefore, it becomes

\begin{equation}
\Delta =\left[ \beta ^{2}+2\gamma ^{2}+k\right] ^{2}-4(\gamma ^{2}+k)(\beta
^{2}+\gamma ^{2}-\epsilon ^{2})=0.
\end{equation}
Solving Eq.(20) for the constant $k,$ we obtain the double roots as $%
k_{1,2}=\beta ^{2}-2\epsilon ^{2}\pm 2\epsilon b,$ where $b=\sqrt{\epsilon
^{2}-\text{\ }\beta ^{2}-\gamma ^{2}}=\frac{1}{2}\left[ \sqrt{1-4\gamma ^{2}}%
-(2n+1)\right] -\epsilon $ with $n=0,1,\cdots .$ Thus, substituting these
values for each $k$ into Eq.(19), we obtain

\begin{equation}
\pi (s)=\pm \left\{
\begin{array}{c}
\left( b-\epsilon \right) s-b;\text{ \ \ \ for \ \ }k_{1}=\beta
^{2}-2\epsilon ^{2}+2\epsilon b, \\
\left( b+\epsilon \right) s-b;\text{ \ \ for \ \ }k_{2}=\beta ^{2}-2\epsilon
^{2}-2\epsilon b.
\end{array}
\right.
\end{equation}
Hence, making the following choice for the polynomial $\pi (s)$ as

\begin{equation}
\pi (s)=-\left( b+\epsilon \right) s+b,
\end{equation}
for $k_{2}=\beta ^{2}-2\epsilon ^{2}-2\epsilon b,$ giving the function:

\begin{equation}
\tau \text{(s)}=-2(1+b+\epsilon )s+1+2b,
\end{equation}
which has a \ negative derivative of the form $\tau
{\acute{}}%
(s)=-2(1+b+\epsilon )=2n-1-\sqrt{1-4\gamma ^{2}}.$ Thus, from Eqs.(8)-(9)
and Eqs.(22)-(23), we find

\begin{equation}
\lambda =-\gamma ^{2}-(b+\epsilon )(b+\epsilon +1),
\end{equation}
and

\begin{equation}
\lambda _{n}=n^{2}+n+2n\left( \epsilon +b\right) .
\end{equation}
Therefore, after setting $\lambda _{n}=\lambda $ and solving for $E_{nq},$
we find the KG exact binding energy spectra as

\begin{equation}
E_{nq}=-\frac{V_{0}}{2q}\pm \xi \sqrt{\frac{m^{2}}{4V_{0}^{2}+\xi ^{2}}-%
\frac{1}{16q^{2}}},
\end{equation}
where

\begin{equation}
\xi =\sqrt{q^{2}\alpha ^{2}-4V_{0}^{2}}-q\alpha (2n+1),
\end{equation}
providing that the condition $q^{2}\alpha ^{2}\geq 4V_{0}^{2}$ must be
fulfilled for any possible bound-states. In this context, it is worthwhile
to point out that \c{S}im\c{s}ek and E\u{g}rifes [31] have recently obtained
a similar expression for the Hulth\'{e}n potential. We find the
corresponding wavefunctions by applying the ${\rm NU}$ method to find the
hypergeometric function $y_{n}(s)$ which is the polynomial solution of
hypergeometric-type equation (3) described with the weight function [42]. By
substituting $\pi (s)$ and $\sigma (s)$ in Eq.(6) and then solving the
first-order differential equation, we find

\begin{equation}
\phi _{n}(s)=s^{b}(1-s)^{\epsilon }.\text{ }
\end{equation}
To find the function $y_{nq}(s),$ which is the polynomial solution of
hypergeometric-type equation, we multiply Eq.(3) by $\rho (s)$ so that it
can be written in self-adjoint form [42]

\begin{equation}
(\sigma (s)\rho (s)y_{nq}^{\prime }(s))^{\prime }+\lambda \rho
(s)y_{nq}(s)=0,
\end{equation}
where $\rho (s)$ satisfies the differential equation $(\sigma (s)\rho
(s))^{\prime }=\tau (s)\rho (s)$ which gives

\begin{equation}
\rho (s)=s^{2b}(1-s)^{2\epsilon }.
\end{equation}
We then obtain the eigenfunction of hypergeometric-type equation from the
Rodrigues relation given by Eq.(4) as

\begin{equation}
y_{nq}(s)=D_{nq}s^{-2b}(1-s)^{-2\epsilon }\frac{d^{n}}{ds^{n}}\left[
s^{n+2b}\left( 1-s\right) ^{n+2\epsilon }\right] ,
\end{equation}
where $D_{nq}$ is a normalizing constant$.$ In the limit $q\rightarrow 1,$
the polynomial solutions of $\ y_{n}(s)$ are expressed in terms of Jacobi
Polynomials, which is one of the classical orthogonal polynomials, with
weight function (30) in the closed interval $\left[ 0,1\right] ,$ giving $%
y_{n,1}(s)\simeq P_{n}^{(2b,2\epsilon )}(1-2s)$ [42]. The radial wave
function $\psi _{nq}(s)$ is obtained from the Jacobi polynomials in Eq.(31)
and $\phi (s)$ in Eq.(28) for the ${\rm s}$-wave functions could be
determined as

\begin{equation}
\psi _{nq}(s)=N_{nq}s^{-b}(1-s)^{-\epsilon }\frac{d^{n}}{ds^{n}}\left[
s^{n+2b}\left( 1-s\right) ^{n+2\epsilon }\right] =N_{nq}s^{b}(1-s)^{\epsilon
}P_{n}^{(2b,2\epsilon )}(1-2s),
\end{equation}
where $b=\frac{\xi }{2q\alpha }-\frac{\sqrt{m^{2}-E_{n}^{2}}}{\alpha },$ $%
s=(1+qe^{-\alpha x})^{-1}$ and $N_{nq}$ is a new normalization constant.

We now make use of the fact that the Jacobi polynomials can be explicitly
written in two different ways [48]:

\begin{equation}
P_{n}^{(\rho ,\nu )}(z)=2^{-n}\sum\limits_{p=0}^{n}(-1)^{n-p}%
{n+\rho  \choose p}%
{n+\nu  \choose n-p}%
\left( 1-z\right) ^{n-p}\left( 1+z\right) ^{p},
\end{equation}

\begin{equation}
P_{n}^{(\rho ,\nu )}(z)=\frac{\Gamma (n+\rho +1)}{n!\Gamma (n+\rho +\nu +1)}%
\sum\limits_{r=0}^{n}%
{n \choose r}%
\frac{\Gamma (n+\rho +\nu +r+1)}{\Gamma (r+\rho +1)}\left( \frac{z-1}{2}%
\right) ^{r},
\end{equation}
where $%
{n \choose r}%
=\frac{n!}{r!(n-r)!}=\frac{\Gamma (n+1)}{\Gamma (r+1)\Gamma (n-r+1)}.$ Using
Eqs.(33)-(34), we obtain the explicit expressions for $P_{n}^{(2b,2\epsilon
)}(1-2s)$

\[
P_{n}^{(2b,2\epsilon )}(1-2s)=(-1)^{n}\Gamma (n+2b+1)\Gamma (n+2\epsilon +1)
\]

\begin{equation}
\times \sum\limits_{p=0}^{n}\frac{(-1)^{p}q^{n-p}}{p!(n-p)!\Gamma
(p+2\epsilon +1)\Gamma (n+2b-p+1)}s^{n-p}(1-s)^{p},
\end{equation}

\begin{equation}
P_{n}^{(2b,2\epsilon )}(1-2s)=\frac{\Gamma (n+2b+1)}{\Gamma (n+2b+2\epsilon
+1)}\sum\limits_{r=0}^{n}\frac{(-1)^{r}q^{r}\Gamma (n+2b+2\epsilon +r+1)}{%
r!(n-r)!\Gamma (2b+r+1)}s^{r}.
\end{equation}

\[
1=N_{nq}^{2}(-1)^{n}\frac{\Gamma (n+2\epsilon +1)\Gamma (n+2b+1)^{2}}{\Gamma
(n+2\epsilon +2b+1)}\left\{ \sum\limits_{p=0}^{n}\frac{(-1)^{p}q^{n-p}}{%
p!(n-p)!\Gamma (p+2\epsilon +1)\Gamma (n+2b-p+1)}\right\}
\]

\begin{equation}
\times \left\{ \sum\limits_{r=0}^{n}\frac{(-1)^{r}q^{r}\Gamma (n+2\epsilon
+2b+r+1)}{r!(n-r)!\Gamma (2b+r+1)}\right\} I_{nq}(p,r),
\end{equation}
where

\begin{equation}
I_{nq}(p,r)=\int\limits_{0}^{1}s^{n+2b+r-p}(1-s)^{p+2\epsilon }ds.
\end{equation}
Using the following integral representation of the hypergeometric function
[49]

\[
\int\limits_{0}^{1}s^{\alpha _{0}-1}(1-s)^{\gamma _{0}-\alpha
_{0}-1}(1-qs)^{-\beta _{0}}ds=_{2}F_{1}(\alpha _{0},\beta _{0}:\gamma _{0};q)%
\frac{\Gamma (\alpha _{0})\Gamma (\gamma _{0}-\alpha _{0})}{\Gamma (\gamma
_{0})},
\]

\begin{equation}
\lbrack
\mathop{\rm Re}%
(\gamma _{0})>%
\mathop{\rm Re}%
(\alpha _{0})>0,\text{ \ }\left| \arg (1-q)\right| <\pi ]
\end{equation}
which gives

\begin{equation}
_{2}F_{1}(\alpha _{0},\beta _{0}:\alpha _{0}+1;q)/\alpha
_{0}=\int\limits_{0}^{1}s^{\alpha _{0}-1}(1-qs)^{-\beta _{0}}ds,
\end{equation}

\[
_{2}F_{1}(\alpha _{0},\beta _{0}:\gamma _{0};q)=\frac{\Gamma (\gamma
_{0})\Gamma (\gamma _{0}-\alpha _{0}-\beta _{0})}{\Gamma (\gamma _{0}-\alpha
_{0})\Gamma (\gamma _{0}-\beta _{0})},
\]
\begin{equation}
\lbrack
\mathop{\rm Re}%
(\gamma _{0}-\alpha _{0}-\beta _{0})>0,\text{ }%
\mathop{\rm Re}%
(\gamma _{0})>%
\mathop{\rm Re}%
(\beta _{0})>0,
\end{equation}
for $q=1.$ Setting $\alpha _{0}=n+2b+r-p+1,$ $\beta _{0}=-p-2\epsilon ,$ and
$\gamma _{0}=\alpha _{0}+1,$ one gets

\begin{equation}
I_{nq}(p,r)=\frac{_{2}F_{1}(\alpha _{0},\beta _{0}:\gamma _{0};q)}{\alpha
_{0}}=\frac{(n+2b+r-p+1)!(p+2\epsilon )!}{(n+2b+r-p+1)(n+2\epsilon +r+2b+1)!}
\end{equation}

\subsection{Real Potentials}

Consider the parameters $V_{0},q,$ and $\alpha $ given in Eq.(13) are all
real, then

(i) For any given $\alpha $ the spectrum consists of real eigenstate spectra
$E_{n}(V_{0},q,\alpha )$ depending on $q.$ The sign of $V_{0}$ does not
affect the bound states. For positive shape parameter $q,$ it is clear that
while $V_{0}\rightarrow 0,$ $E_{n}=m\sqrt{1-\left( \frac{\alpha n}{2m}%
\right) ^{2}}$ tends to the value $m$ for the ground state (i.e., $n=0)$ and
$0.866m$ for the first excited state (i.e., $n=1),$ etc. In these
calculations, we have used $\lambda _{c}=\frac{\hbar }{mc}=\frac{1}{m}=\frac{%
1}{\alpha \text{ }}$ which is the compton wavelength of the KG particles.

(ii) There exist bound states (real solution) in case if the condition $%
4V_{0}^{2}\leq q^{2}\alpha ^{2}$ is achieved, otherwise there are no
bound-states.

(iii) There exist bound states in case if the condition $4V_{0}^{2}+\xi
^{2}\leq 16q^{2}m^{2}$ is achieved, otherwise there are no bound-states.

Moreover, this condition which gives the critical coupling value turns to be

\begin{equation}
n\leq \frac{1}{q\alpha }\left( \sqrt{4q^{2}m^{2}-V_{0}^{2}}+\sqrt{\frac{%
q^{2}\alpha ^{2}}{4}-V_{0}^{2}}\right) -\frac{1}{2},
\end{equation}
i.e., there are only finitely many eigenstates. In order that at least one
level might exist $(n=0)$, its necessary that the inequality

\begin{equation}
q\alpha -\sqrt{q^{2}\alpha ^{2}-4V_{0}^{2}}\leq 2\sqrt{4q^{2}m^{2}-V_{0}^{2}}%
,
\end{equation}
is fulfilled

\subsection{Complex Potentials}

Under ${\cal P}$, the spatial coordinates $(x,y,z)$ are replaced by $%
(-x,-y,-z)$ but $r$ is replaced by $r$ and not $-r,$ in the radial wave
equation (4). Thus, the $s$-wave differential equation is not ${\cal PT}{\rm %
-}$symmetric. The radial Schr\"{o}dinger wave equation becomes a different
differential equation under the action of the ${\cal PT}{\rm -}$operator and
does not go into itself. This means that we must solve the problem in $1D,$
on the full plane, say $x$-direction and not in the radial direction $r.$

\subsubsection{Non-Hermitian ${\cal PT}{\rm -S}$ymmetric Generalized
Woods-Saxon Potential}

Let us consider the case where at least one of the potential parameters be
complex:

If $\alpha $ is a complex parameter ($\alpha \rightarrow i\alpha $), the
potential (13) becomes

\begin{equation}
V_{q}(x)=-\frac{V_{0}}{q^{2}+2q\cos (\alpha x)+1}\left[ q+\cos (\alpha
x)-i\sin (\alpha x)\right] =V_{q}^{\ast }(-x),
\end{equation}
which is a ${\cal PT}$-symmetric but non-Hermitian. It has real spectrum
given by

\begin{equation}
E_{nq}=-\frac{V_{0}}{2q}+\left( \sqrt{q^{2}\alpha ^{2}+4V_{0}^{2}}-\alpha
q(2n+1)\right) \sqrt{\frac{1}{16q^{2}}-\frac{m^{2}}{4V_{0}^{2}-\left( \sqrt{%
q^{2}\alpha ^{2}+4V_{0}^{2}}-q\alpha (2n+1)\right) ^{2}}},
\end{equation}
if and only if $16q^{2}m^{2}\leq 4V_{0}^{2}-\left( \sqrt{q^{2}\alpha
^{2}+4V_{0}^{2}}-q\alpha (2n+1)\right) ^{2}\ $\ which gives $\frac{8qm^{2}}{%
\alpha (2n+1)}+\frac{q\alpha }{2(2n+1)}+\frac{q\alpha }{2}(2n+1)\leq \sqrt{%
q^{2}\alpha ^{2}+4V_{0}^{2}}.$

Figs. 1(a) and (b) show the variation of the ground-state (i.e., $n=0$) as a
function of the coupling constant $V_{0}$ for different positive and
negative $q$, and $a=\lambda _{c}.$ Obviously, in Fig. 1(a), the
non-Hermitian ${\cal PT}$-symmetric generalized WS potential generates real
and negative bound-states for $q>0,$ it generates real and positive
bound-states for the same value of $\alpha $ when $q<0$ (Fig. 1(b)).
Further, Figs. 2(a) and (b) show the variation of the first three energy
eigenstates as a function of $\alpha $ for (a) $q=1.0,V_{0}=6m$ and (b) $%
q=-1.0,V_{0}=2m.$ Obviously, for the given $V_{0},$ as seen from Figs. 2(a)
and (b) all possible eigenstates have negative (positive) eigenenergies if
the parameter $q$ is positive (negative). It is almost notable that there
are some crossing points of the relativistic energy eigenvalues for some $%
V_{0}$ values.

The corresponding radial wave function $\psi _{nq}(s)$ for the ${\rm s}$%
-wave could be determined as

\begin{equation}
\psi _{nq}(s)=N_{nq}s^{ib}(1-qs)^{i\epsilon }P_{n}^{(2ib,2i\epsilon )}(1-2s),
\end{equation}
where $s=(1+qe^{-i\alpha x})^{-1}.$

For the sake of comparing the relativistic and non-relativistic binding
energies, we need to solve the $1D$ \ Schr\"{o}dinger equation for the
complex form of the generalized WS potential given by Eq.(45). We employ the
convenient transformation $s(x)=(1+qe^{-i\alpha x})^{-1}$, $0\leq r\leq
\infty \rightarrow 0\leq s\leq 1,$ to obtain

\begin{equation}
\psi {}_{nq}^{\prime \prime }(s)+\frac{1-2s}{(s-s^{2})}\psi {}_{nq}^{\prime
}(s)+\frac{\left[ -\beta ^{2}s+\beta ^{2}-\epsilon ^{2}\right] }{\left(
s-s^{2}\right) ^{2}}\psi _{nq}(s)=0,
\end{equation}
for which
\[
\widetilde{\tau }(s)=1-2s,\text{ \ \ \ }\sigma (s)=s-s^{2},\text{ \ \ }%
\widetilde{\sigma }(s)=-\beta ^{2}s+\beta ^{2}-\epsilon ^{2},
\]

\begin{equation}
\epsilon ^{2}=\frac{2m}{\hbar ^{2}\alpha ^{2}}E_{n}\text{ \ (}E_{n}<0),\text{
\ }\beta ^{2}=-\frac{2m}{\hbar ^{2}\alpha ^{2}q}V_{0}\text{ \ \ (}\beta
^{2}>0).
\end{equation}
Moreover, it could be obtained

\begin{equation}
\tau (s)=-2(1+c+\epsilon )s+(1+2c),\text{ c}=\sqrt{\epsilon ^{2}-\text{\ }%
\beta ^{2}},
\end{equation}
if $\pi (s)=-(c+\epsilon )s+c$ is chosen for $k_{-}=-(c+\epsilon )^{2}.$ We
also find the eigenvalues

\begin{equation}
\lambda =-\left( c+\epsilon \right) (c+\epsilon +1),\text{ \ }\lambda
_{n}=2n\left( c+\epsilon +1\right) +n(n-1).
\end{equation}
Finally, setting $\lambda $=$\lambda _{n}$ and solving for $\epsilon ,$ then
the energy eigenvalues of the system under consideration could be found as

\begin{equation}
E_{n}(V_{0},q,i\alpha )=\frac{\hbar ^{2}\alpha ^{2}}{2m}\left[ \frac{n+1}{2}-%
\frac{\gamma }{(n+1)}\right] ^{2},\text{ \ \ }\gamma =\frac{mV_{0}}{\hbar
^{2}q\alpha ^{2}},\text{ \ }0\leq n<\infty .\text{ \ \ }
\end{equation}
On the other hand, the radial wave function in the present case becomes

\begin{equation}
\psi _{nq}(s)=N_{nq}s^{c}(1-s)^{\epsilon }P_{n}^{(2c,2\epsilon )}(1-2s),
\end{equation}
with $s(x)=(1+e^{-i\alpha x})^{-1}$and $N_{nq}$ is a new normalization
constant determine by

\[
1=N_{nq}^{2}(-1)^{n}\frac{(n+2\epsilon )!\Gamma (n+2c+1)^{2}}{\Gamma
(n+2c+2\epsilon +1)}\left\{ \sum\limits_{p=0}^{n}\frac{(-1)^{p}q^{n-p}}{%
p!(n-p)!(2\epsilon +p)!\Gamma (n+2c-p+1)}\right\}
\]

\begin{equation}
\times \left\{ \sum\limits_{r=0}^{n}\frac{(-1)^{r}q^{r}\Gamma
(n+2c+r+2\epsilon +1)}{r!(n-r)!\Gamma (2c+r+1)}\right\}
\int\limits_{0}^{1}s^{n+2c+r-p}(1-qs)^{p+2\epsilon }ds,
\end{equation}
where

\begin{equation}
\int\limits_{0}^{1}s^{n+2c+r-p}(1-qs)^{p+2\epsilon
}ds=_{2}F_{1}(n+2c+r-p+1,-p-2\epsilon :n+2c+r-p+2:1)B(n+2c+r-p+1,1),
\end{equation}

\subsubsection{Non-Hermitian non-${\cal PT}{\rm -S}$ymmetric Generalized
Woods-Saxon Potential}

Let two parameters; namely, $V_{0}$ and $q$ be complex parameters (i.e., $%
V_{0}\rightarrow iV_{0},$ $q\rightarrow iq$), then we obtain the potential as

\begin{equation}
V_{q}(x)=V_{0}\frac{\left[ 2\cosh ^{2}(\alpha x)-\sinh (2\alpha x)-1\right]
-i\left[ \cosh (\alpha x)-\sinh (\alpha x)\right] }{1+q^{2}\left[ 2\cosh
^{2}(\alpha x)-\sinh (2\alpha x)-1\right] }.
\end{equation}
This potential is a non ${\cal PT}$-symmetric but non-Hermitian possesses
exact real spectra

\begin{equation}
E_{nq}=-\frac{V_{0}}{2q}+\left( \sqrt{q^{2}\alpha ^{2}-4V_{0}^{2}}-q\alpha
(2n+1)\right) \sqrt{\frac{m^{2}}{4V_{0}^{2}+\left( \sqrt{q^{2}\alpha
^{2}-4V_{0}^{2}}-q\alpha (2n+1)\right) ^{2}}-\frac{1}{16q^{2}}},
\end{equation}
if and only if $16q^{2}m^{2}\geq 4V_{0}^{2}+\left( \sqrt{q^{2}\alpha
^{2}-4V_{0}^{2}}-q\alpha (2n+1)\right) ^{2}$ which gives $-\frac{8qm^{2}}{%
\alpha (2n+1)}+\frac{q\alpha }{2(2n+1)}+\frac{q\alpha }{2}(2n+1)\leq \sqrt{%
q^{2}\alpha ^{2}-4V_{0}^{2}}.$

On the other hand, the corresponding radial wave functions $\psi _{nq}(s)$
for the ${\rm s}$-wave could be determined as

\begin{equation}
\psi _{nq}(s)=N_{nq}s^{b}(1-s)^{\epsilon }P_{n}^{(2b,2\epsilon )}(1-2s),
\end{equation}
where $s=(1+iqe^{-\alpha x})^{-1}.$ The integral $I_{nq}(p,r)=\int%
\limits_{0}^{1}s^{n+2b+r-p}(1-s)^{p+2\epsilon }ds$ is given by

\begin{equation}
I_{nq}(p,r)=_{2}F_{1}(n+2b+r-p+1,-p-2\epsilon :n+2b+r-p+2;i)B(n+2b+r-p+1,1).
\end{equation}

\subsubsection{Pseudo-Hermiticity and ${\cal PT}{\rm -S}$ymmetric
Generalized Woods-Saxon Potential}

When all the parameters $V_{0},$ $\alpha $ and $q$ are complex parameters
(i.e., $V_{0}\rightarrow iV_{0},$ $\alpha \rightarrow i\alpha ,$ $%
q\rightarrow iq$), we obtain

\begin{equation}
V_{q}(x)=-\frac{V_{0}}{q^{2}+2q\sin (\alpha x)+1}\left[ q+\sin (\alpha
x)+i\cos (\alpha x)\right] =V_{q}^{\ast }(\frac{\pi }{2}-x).
\end{equation}
This potential is a pseudo-Hermitian potential [27,48] having a $\pi /2$
phase difference with respect to the potential (I), $\eta =P$%
-pseudo-Hermitian (i.e., $PTV_{q}(x)(PT)^{-1}=V_{q}(x),$ with $P=\eta
:x\rightarrow \frac{\pi }{2\alpha }-x$ and $T:i\rightarrow -i),$ it is also
a non ${\cal PT}$-symmetric but non-Hermitian having exact real spectrum
given by

\begin{equation}
E_{nq}=-\frac{V_{0}}{2q}+\left( \sqrt{q^{2}\alpha ^{2}+4V_{0}^{2}}+q\alpha
(2n+1)\right) \sqrt{\frac{1}{16q^{2}}-\frac{m^{2}}{4V_{0}^{2}-\left( \sqrt{%
q^{2}\alpha ^{2}+4V_{0}^{2}}+q\alpha (2n+1)\right) ^{2}}},
\end{equation}
if and only if $16q^{2}m^{2}\leq 4V_{0}^{2}-\left( \sqrt{q^{2}\alpha
^{2}+4V_{0}^{2}}+q\alpha (2n+1)\right) ^{2}$ which gives $-\frac{8qm^{2}}{%
\alpha (2n+1)}-\frac{q\alpha }{2(2n+1)}-\frac{q\alpha }{2}(2n+1)\geq \sqrt{%
q^{2}\alpha ^{2}+4V_{0}^{2}}$

Figs. 3(a) and (b) show the variation of the ground-state (i.e., $n=0$) as a
function of the coupling constant $V_{0}$ for different positive $q$ with $%
a=10$ and negative $q$ with $\alpha =1.$ Obviously, in Fig. 3(a), the $P$%
-pseudo-Hermitian non-${\cal PT}$-symmetric generalized WS potential
generates real and negative bound-states for $q>0,$ it also generates real
and positive bound-states for the same value of $\alpha $ when $q<0$ (Fig.
3(b)). Further, Figs. 4(a) and (b) show the variation of the first three
energy eigenstates as a function of $\alpha $ for (a) $q=1.0,V_{0}=2m$ and
(b) $q=-1.0,V_{0}=4m.$ Obviously, for the given $V_{0},$ as seen from
Figs.4(a) and (b) all possible eigenstates have negative (positive)
eigenenergies if the shape parameter $q$ is positive (negative). There are
some limitation on the number of bound states. It is seen that the range
parameter $\alpha $ has limits for any given state.

On the other hand, the corresponding radial wave functions $\psi _{nq}(s)$
for the ${\rm s}$-wave could be determined as

\begin{equation}
\psi _{nq}(s)=N_{nq}s^{ib}(1-s)^{i\epsilon }P_{n}^{(2ib,2i\epsilon )}(1-2s),
\end{equation}
with $s=(1+iqe^{-i\alpha x})^{-1}.$ The integral $I_{nq}(p,r)=\int%
\limits_{0}^{1}s^{n+2ib+r-p}(1-s)^{p+2i\epsilon }ds$ is given by

\begin{equation}
I_{nq}(p,r)=_{2}F_{1}(n+2ib+r-p+1,-p-2i\epsilon
:n+2ib+r-p+2;i)B(n+2ib+r-p+1,1).
\end{equation}

\section{Results And Conclusions}

\label{RAC}We have seen that the $s$-wave KG equation for the generalized WS
potential can be solved exactly. The relativistic bound-state energy
spectrum and the corresponding wave functions for the generalized WS
potential have been obtained by the NU method. Some interesting results
including the ${\cal PT}$-symmetric and pseudo-Hermitian versions of the
generalized WS potential have also been discussed for the real bound-states.
In addition, we have discussed the relation between the non-relativistic and
relativistic solutions and the possibility of existence of bound states for
complex parameters. We have also shown the possibility to obtain
relativistic bound-states of complex quantum mechanical formulations.
Finally, the relativistic model provides real solution for the complex
exponential potential where this solution is not available in the
nonrelativistic model.

\acknowledgments This research was partially supported by the
Scientific and Technological Research Council of Turkey. S.M.
Ikhdair wishes to dedicate this work to his family for their love
and assistance.

\bigskip

\baselineskip= 2\baselineskip
\bigskip

\bigskip

\begin{figure}[tbp]
\caption{The variation of the ground-state $(n=0)$ energy with $\protect%
\alpha =1$, in a non-Hermitian ${\cal PT}$-symmetric potential given by
Eq.(45), as a function of $V_{0}$ for three different (a) positive and (b)
negative $q.$ }
\label{Figure 1}
\end{figure}
\bigskip
\begin{figure}[tbp]
\caption{The variation of the first three energy eigenstates, in a
non-Hermitian ${\cal PT}$-symmetric potential given by Eq.(45), as a
function of $\protect\alpha $ for (a) $q=1.0,V_{0}=6m$ and (b) $%
q=-1.0,V_{0}=2m.$}
\label{Figure 2}
\end{figure}

\begin{figure}[tbp]
\caption{The variation of the ground-state $(n=0)$ energy, in a $P$-pseudo
Hermitian non-${\cal PT}$-symmetric potential given by Eq.(60), as a
function of $V_{0}$ for three different (a) positive $q$ with $a=10$ and (b)
negative $q$ with $\protect\alpha =1.$}
\label{Figure 3}
\end{figure}
\bigskip
\begin{figure}[tbp]
\caption{The variation of the first three energy eigenstates, in a $P$%
-pseudo -Hermitian non-${\cal PT}$-symmetric potential given by Eq.(60), as
a function of $\protect\alpha $ for (a) $q=1.0,V_{0}=2m$ and (b) $%
q=-1.0,V_{0}=4m.$}
\label{Figure4}
\end{figure}

\newpage

\begin{figure}
\epsfig{file=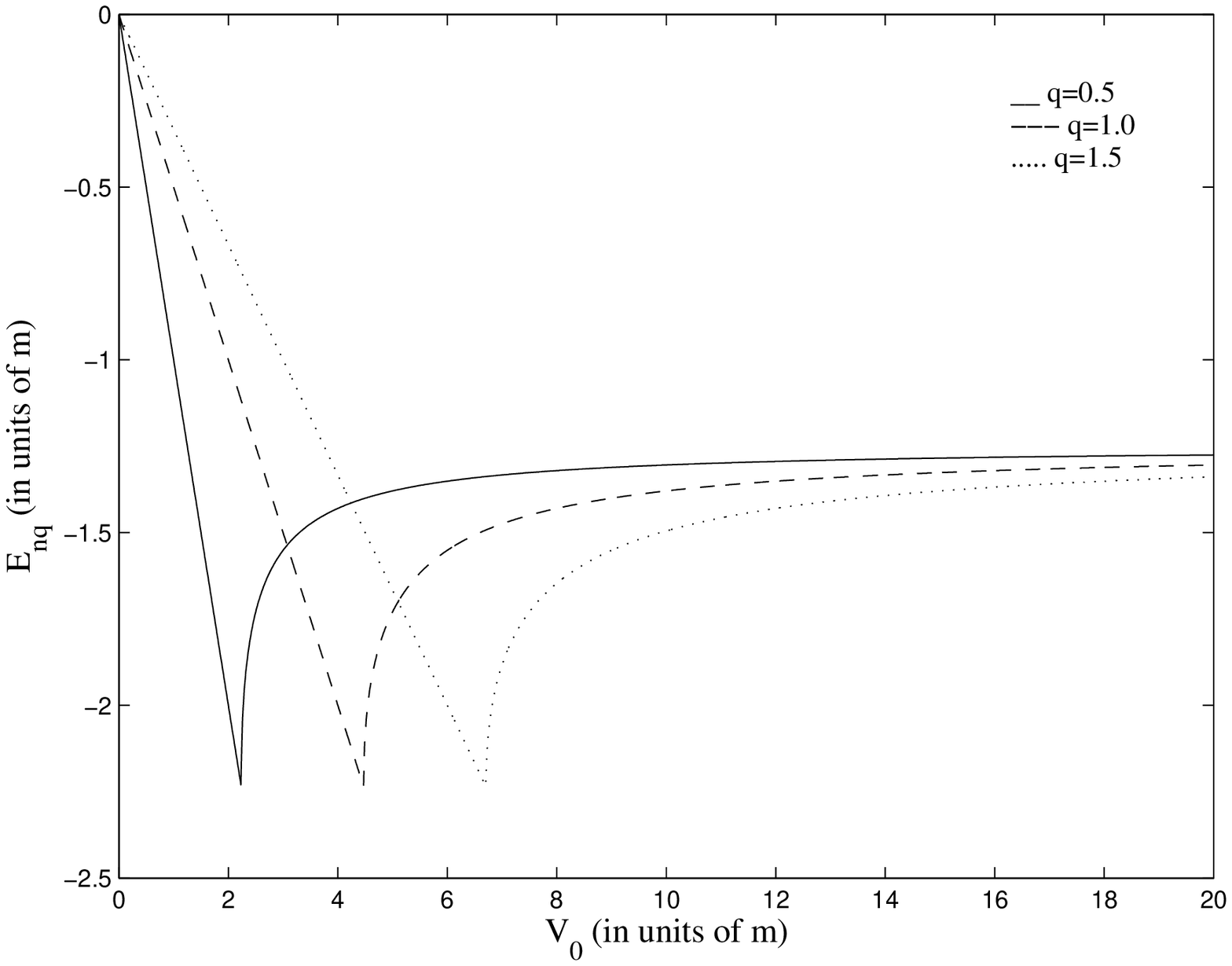,width=17cm,angle=0} \label{Fig1a}
\end{figure}

\newpage

\begin{figure}
\epsfig{file=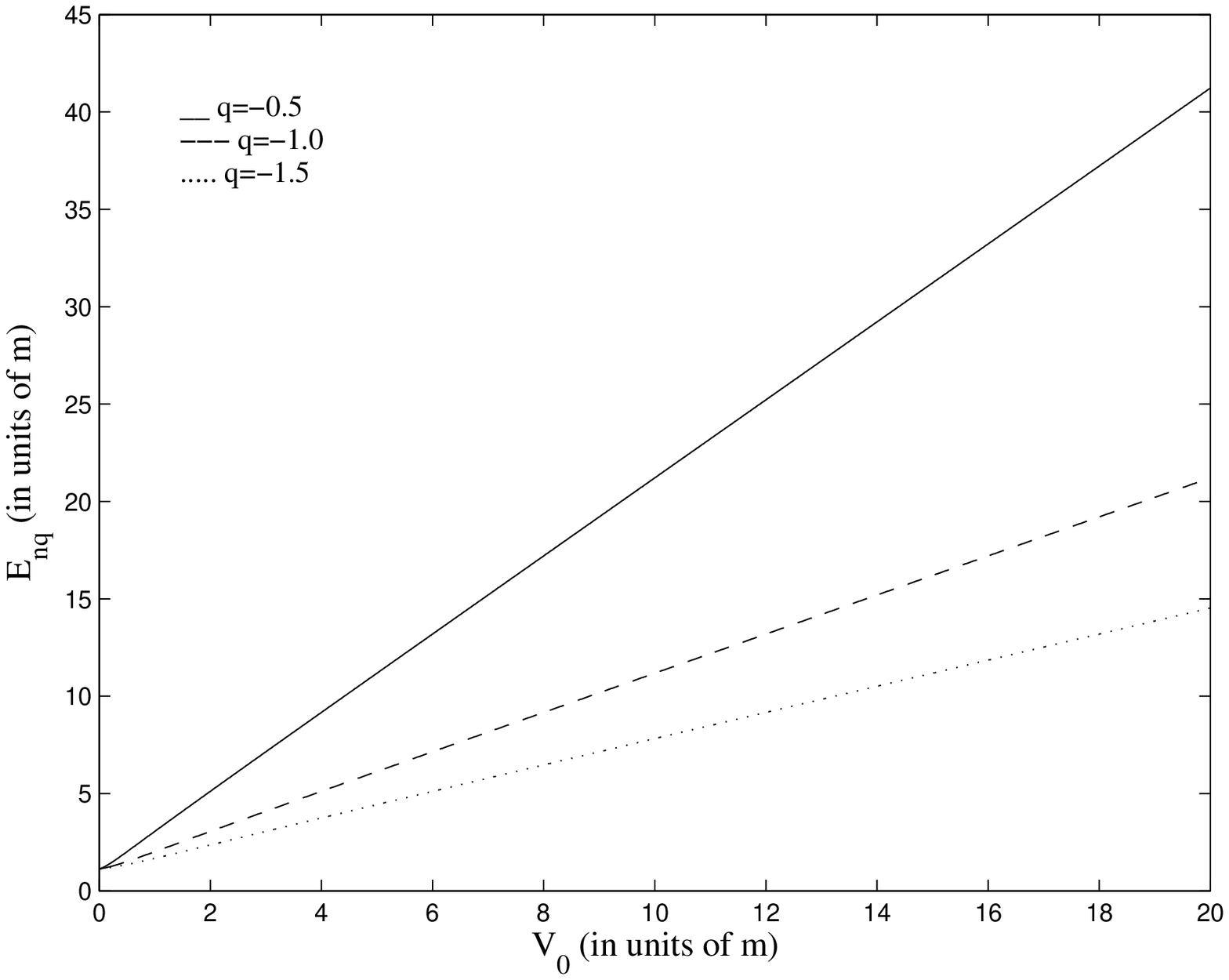,width=17cm,angle=0} \label{Fig1b}
\end{figure}

\newpage

\begin{figure}
\epsfig{file=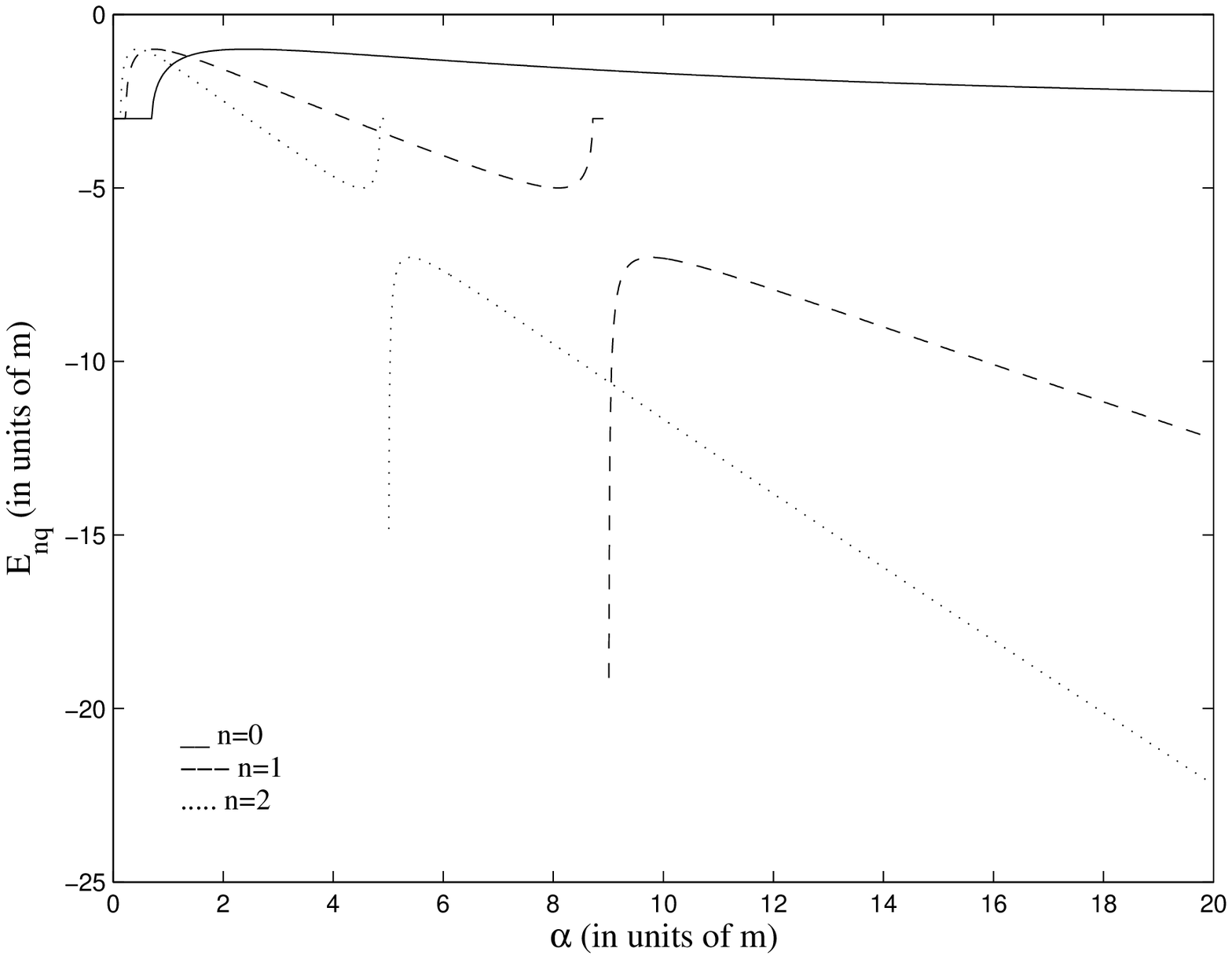,width=17cm,angle=0} \label{Fig2a}
\end{figure}

\newpage

\begin{figure}
\epsfig{file=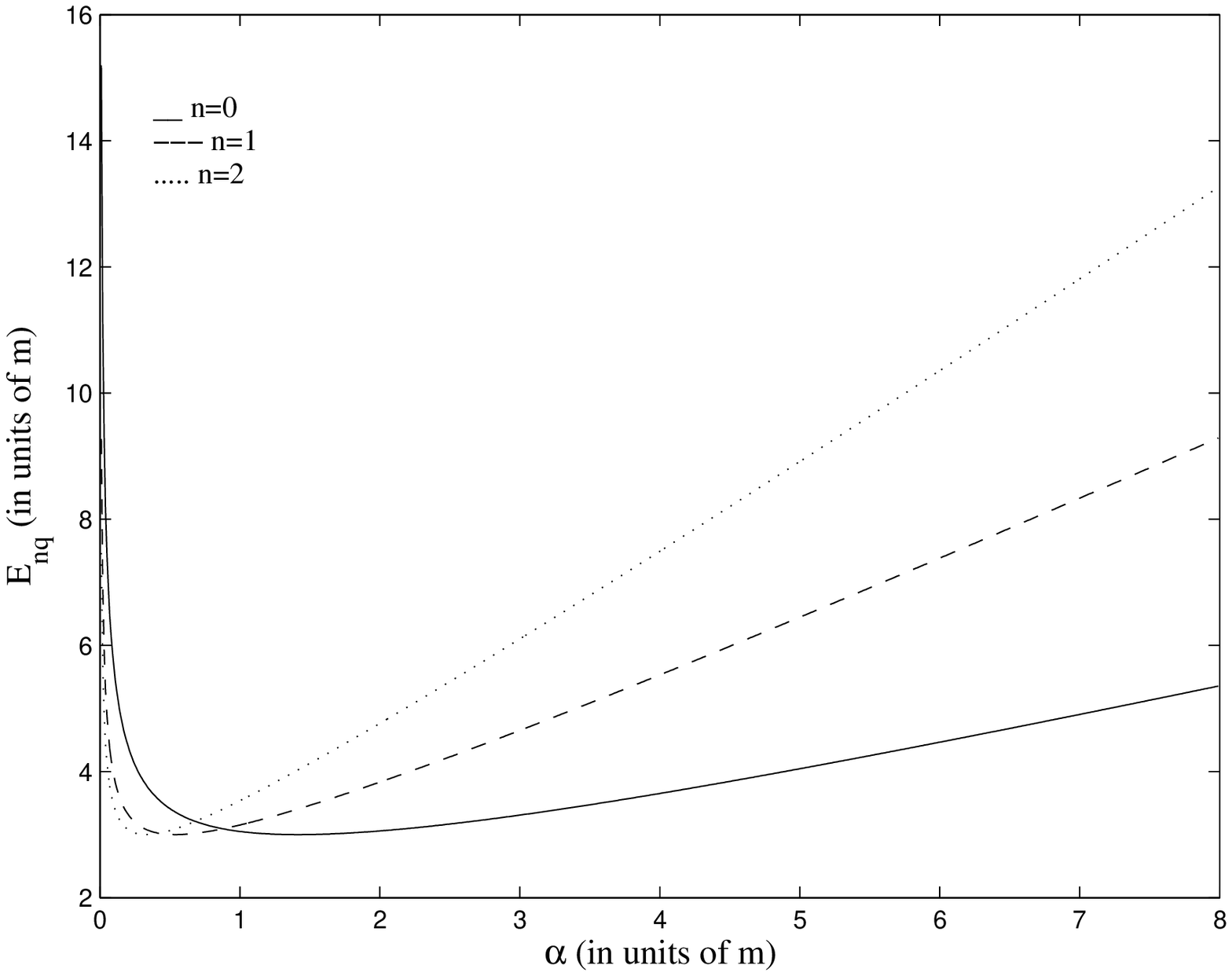,width=17cm,angle=0} \label{Fig2b}
\end{figure}

\newpage

\begin{figure}
\epsfig{file=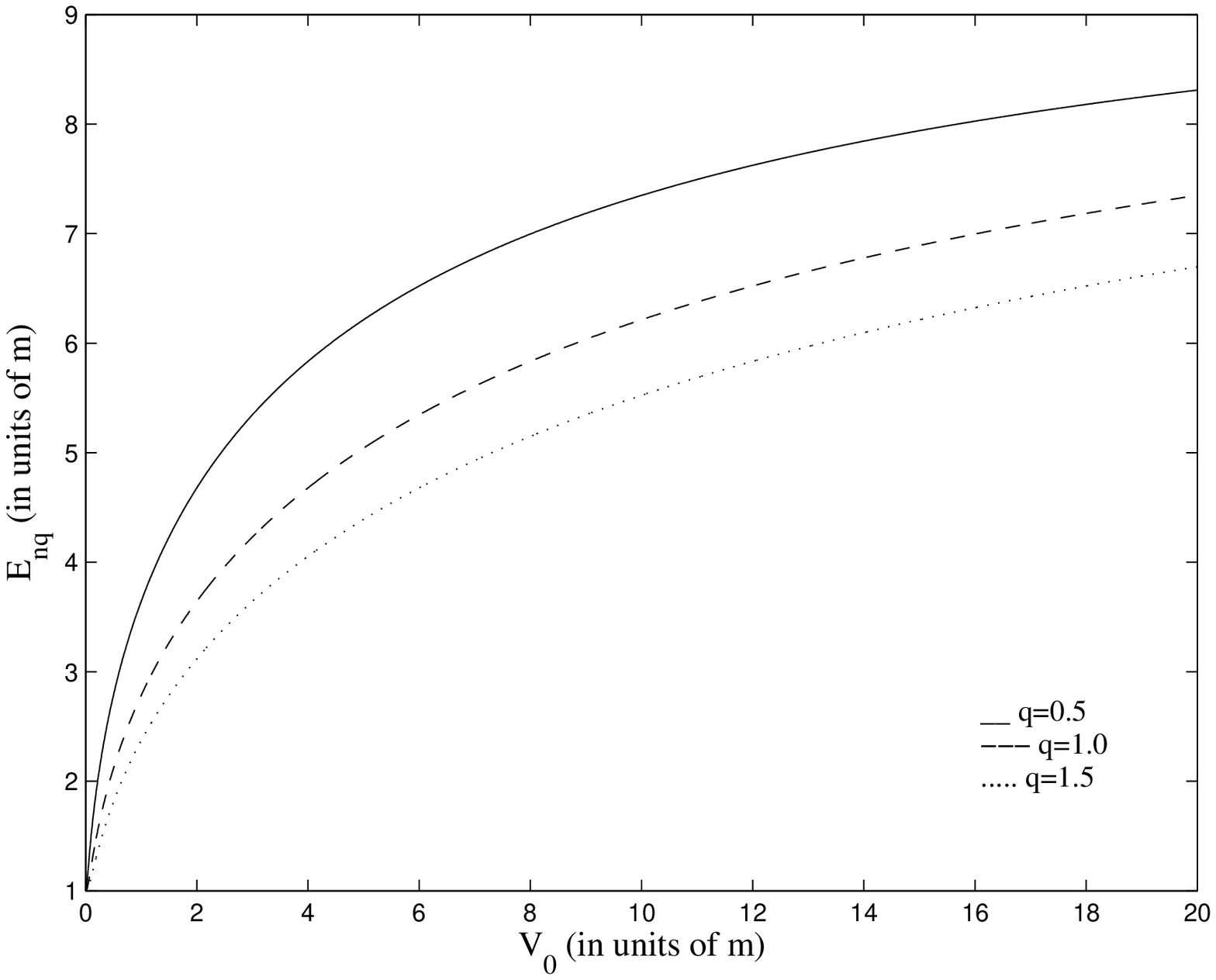,width=17cm,angle=0} \label{Fig3a}
\end{figure}

\newpage

\begin{figure}
\epsfig{file=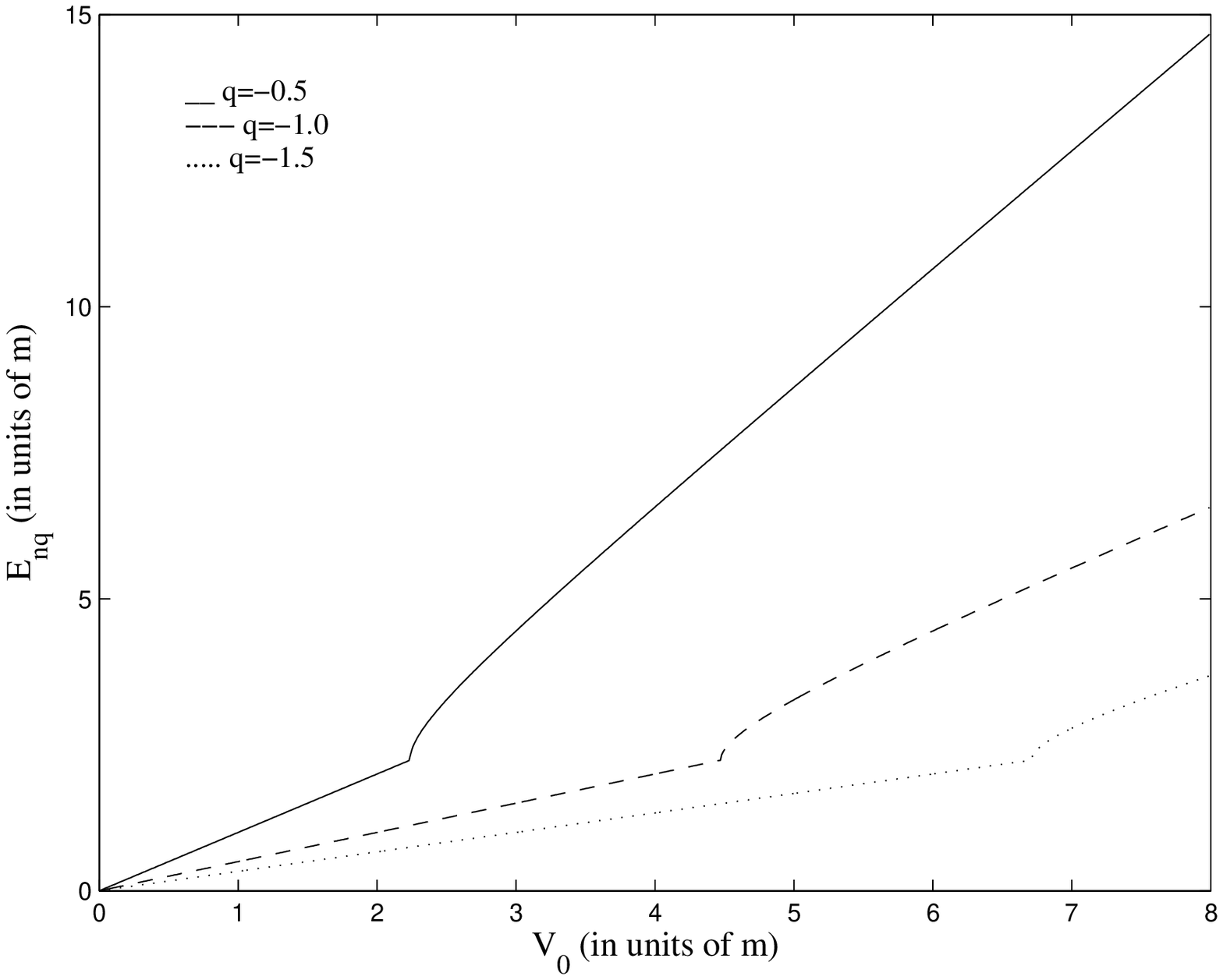,width=17cm,angle=0} \label{Fig3b}
\end{figure}

\newpage

\begin{figure}
\epsfig{file=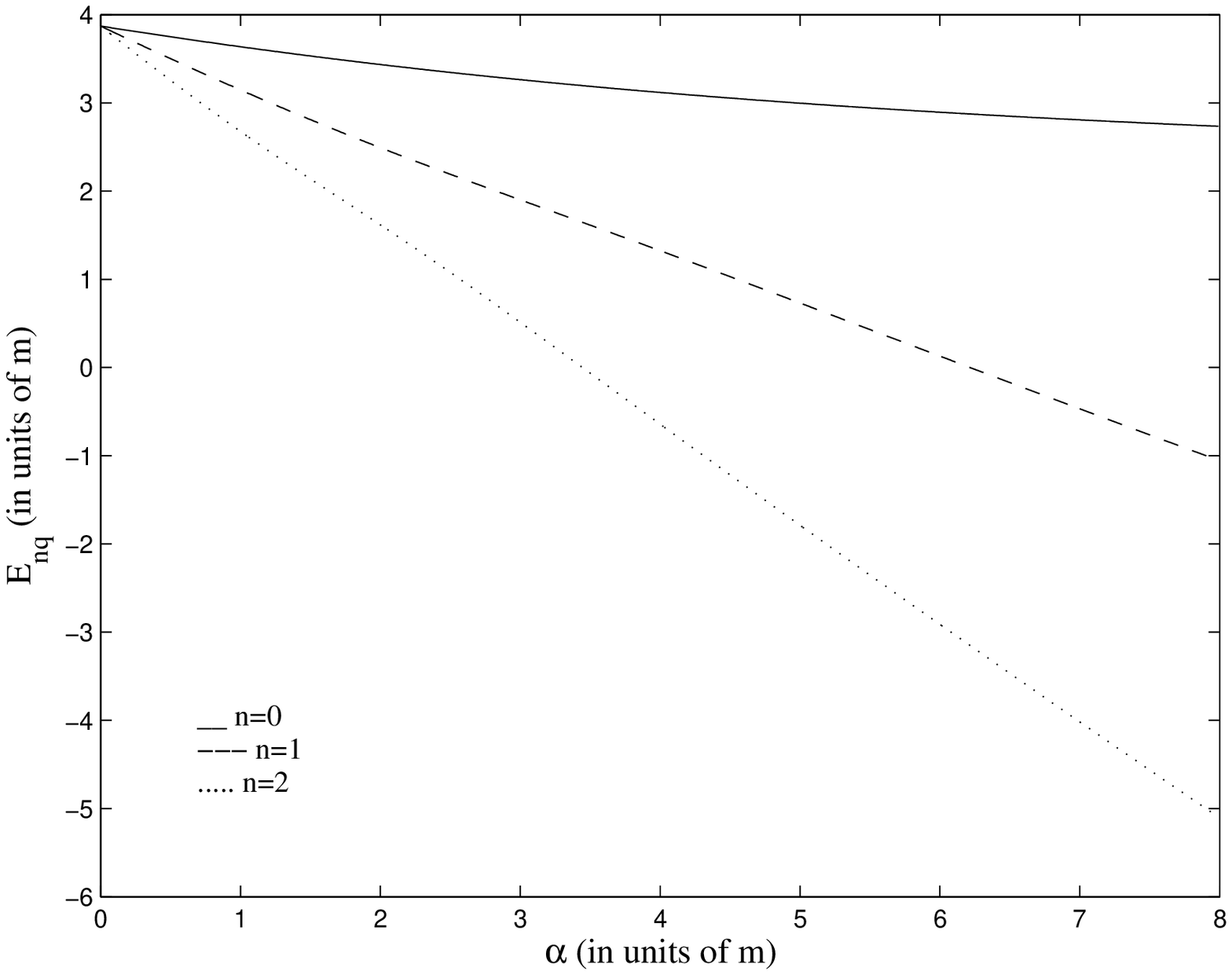,width=17cm,angle=0} \label{Fig4a}
\end{figure}

\newpage

\begin{figure}
\epsfig{file=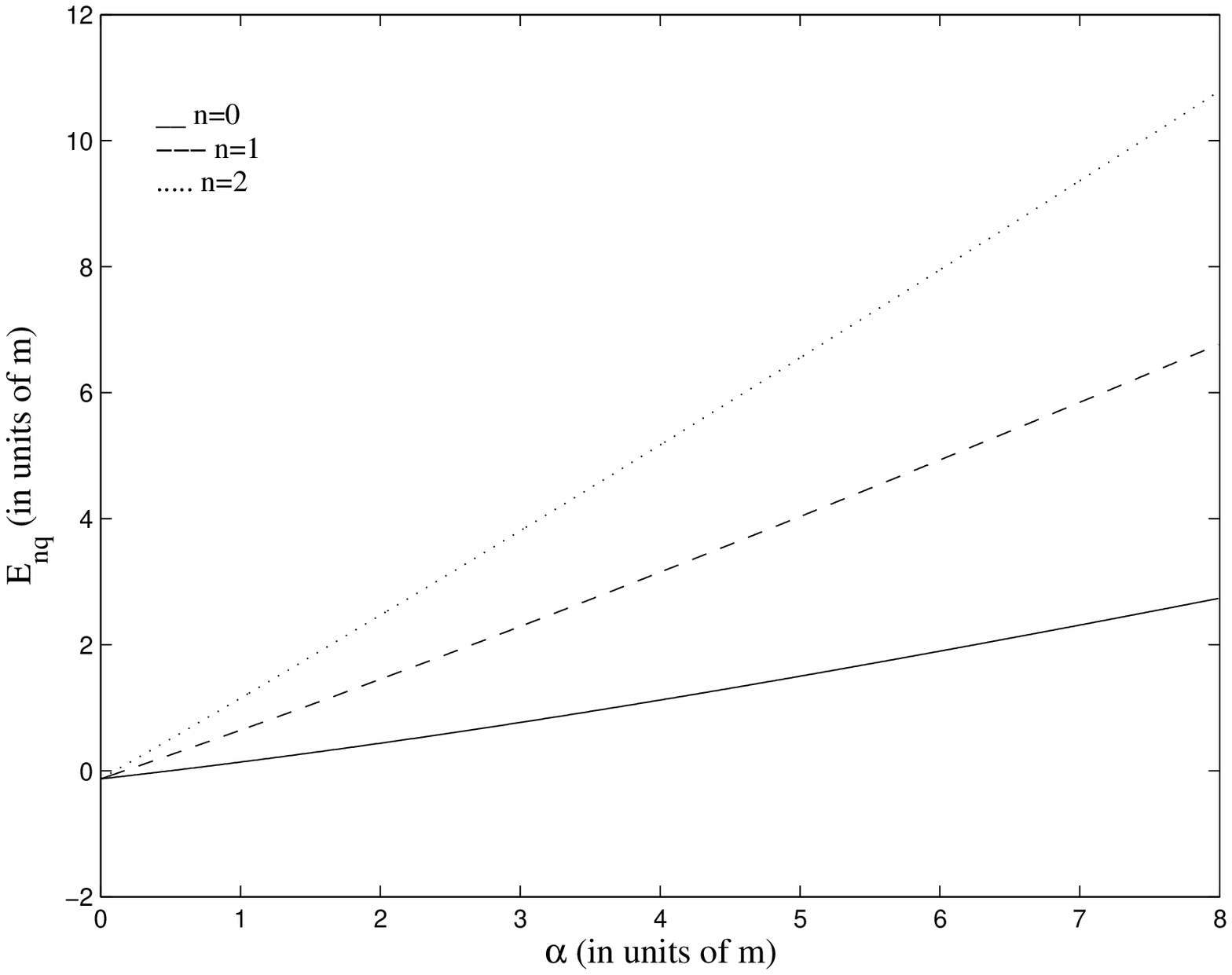,width=17cm,angle=0} \label{Fig4b}
\end{figure}
\end{document}